\documentclass[12pt]{article}
\usepackage{epsfig}

\newlength{\dinwidth} \newlength{\dinmargin}
\begin{document}

\begin{center}
{\Large \bf Large-$Q_T$ $W$-boson production at the 
Tevatron}\footnote{Presented at DPF04, Riverside, California, 
August 26-31, 2004.}
\end{center}
\vspace{2mm}

\begin{center}
{\large Nikolaos Kidonakis$^a$ and
Agust{\' \i}n Sabio Vera$^b$}\\
\vspace{2mm}
{\it $^a$ Cavendish Laboratory, University of Cambridge\\
Madingley Road, Cambridge CB3 0HE, England\\
and\\
Kennesaw State University, 1000 Chastain Rd., \#1202\\
Kennesaw, GA 30144-5591, USA\\
\vspace{2mm}
$^b$ II. Institut f{\"u}r Theoretische Physik, Universit{\"a}t Hamburg\\
Luruper Chaussee 149, 22761~Hamburg, Germany}
\end{center}
 
\vspace{3mm}
 
\begin{abstract}

The production of $W$ bosons at large transverse momentum at the Tevatron
is dominated by soft-gluon corrections. In this talk we present a 
calculation of these corrections at next-to-next-to-leading order.
The corrections enhance the transverse momentum distribution of the $W$
while reducing the scale dependence. 
 
\end{abstract}
 
\thispagestyle{empty} \newpage \setcounter{page}{2}
 
\section{Introduction}
$W$ hadroproduction is useful in estimates of backgrounds
to new physics (such as Higgs production). The transverse momentum, 
$Q_T$, distribution of the $W$ falls rapidly by several orders
of magnitude as $Q_T$ increases.
 
Full next-to-leading order (NLO) results for $W$ hadroproduction
at large $Q_T$ have been available
for some time \cite{AR,gpw}. 
At lowest order the partonic channels involved are
$q(p_a) + g(p_b) \longrightarrow W(Q) + q(p_c)$
and $q(p_a) + {\bar q}(p_b) \longrightarrow W(Q) + g(p_c)$.
We define $s=(p_a+p_b)^2$, $t=(p_a-Q)^2$, $u=(p_b-Q)^2$
and $s_2=s+t+u-Q^2$.
At threshold, i.e. when we have just enough energy to produce
a $W$ with a certain $Q_T$, $s_2 \rightarrow 0$.
 
The large-$Q_T$ distribution is enhanced by 
soft-gluon corrections, which are dominant near threshold.
These corrections are of the form
${\cal D}_l(s_2)\equiv[\ln^l(s_2/Q_T^2)/s_2]_+$.
For the order $\alpha_s^n$ corrections $l\le 2n-1$.
At NLO in $\alpha_s$, we have terms with ${\cal D}_1(s_2)$ and
${\cal D}_0(s_2)$ logarithms,
as well as $\delta(s_2)$ terms that involve the virtual corrections.
 
At next-to-next-to-leading order (NNLO) in $\alpha_s$, 
we have terms with ${\cal D}_3(s_2)$, ${\cal D}_2(s_2)$,
${\cal D}_1(s_2)$, and ${\cal D}_0(s_2)$ logarithms,
as well as $\delta(s_2)$ terms for the virtual corrections. 
Thus, at NNLO,  the leading logs (LL) are ${\cal D}_3(s_2)$,
the next-to-leading logs (NLL) are ${\cal D}_2(s_2)$, the
next-to-next-to-leading logs (NNLL) are ${\cal D}_1(s_2)$, and the
next-to-next-to-next-to-leading logs (NNNLL) are ${\cal D}_0(s_2)$.
 
We can formally resum these soft logarithms to all orders in 
$\alpha_s$ \cite{KS,LOS,NK}. This has been done explicitly for $W$ 
production in Ref.~\cite{NKVD}. However, for numerical results
here we expand the resummed formula to NNLO
to avoid using prescriptions for the resummed cross section \cite{NKRV}. 
  
A unified approach and a master formula
for calculating these soft logarithms at NNLO for any process
has been presented in Ref.~\cite{NKuni}.
It has been  applied to $W$ production in Ref.~\cite{NKASV}.

\section{$W$ production with large $Q_T$ at the Tevatron}

We now present our numerical results for large-$Q_T$ $W$-boson
production \cite{NKASV} at the Fermilab Tevatron.

\begin{figure}
\centerline{\psfig{file=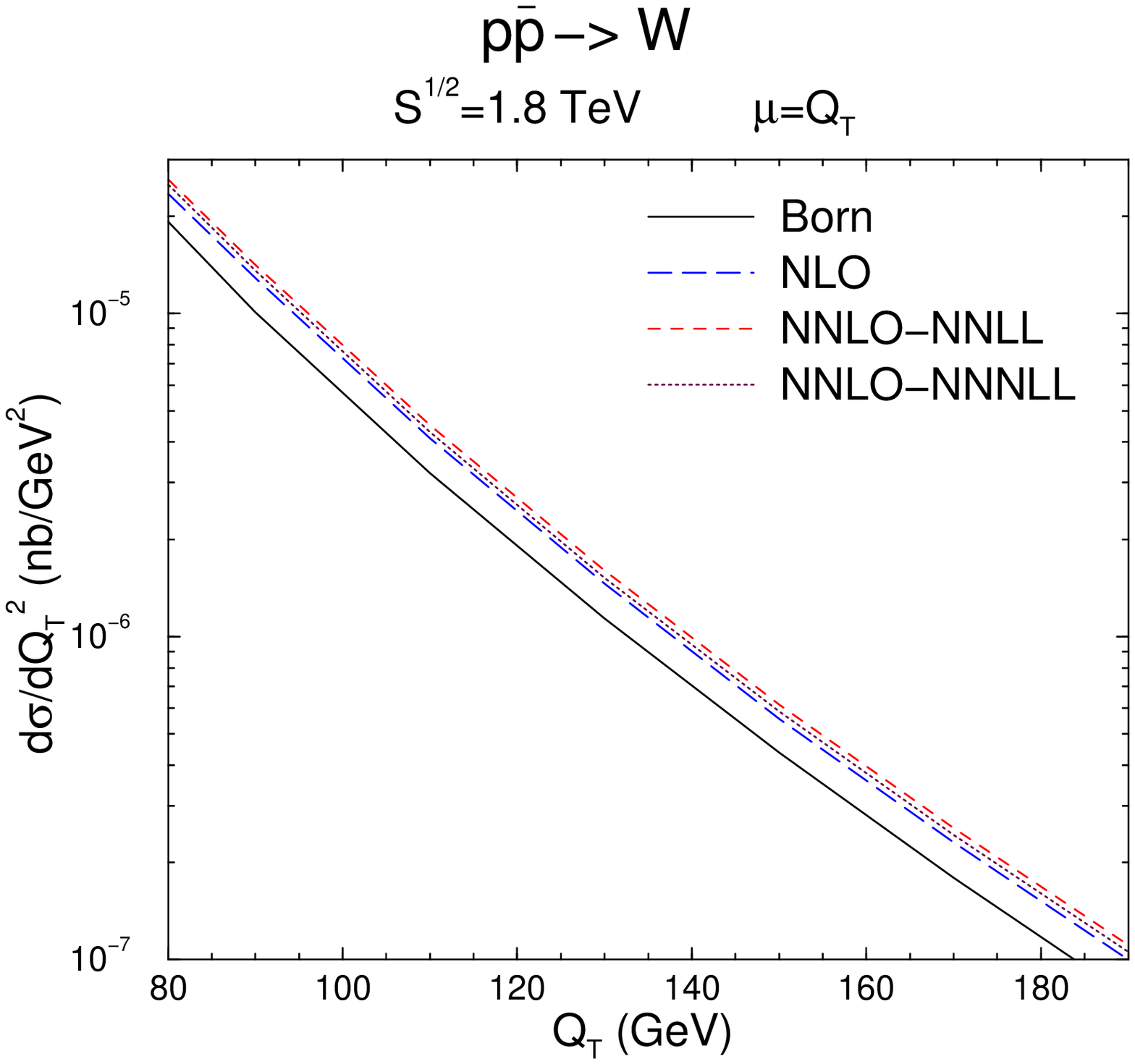,width=8cm} \hspace{5mm}
\psfig{file=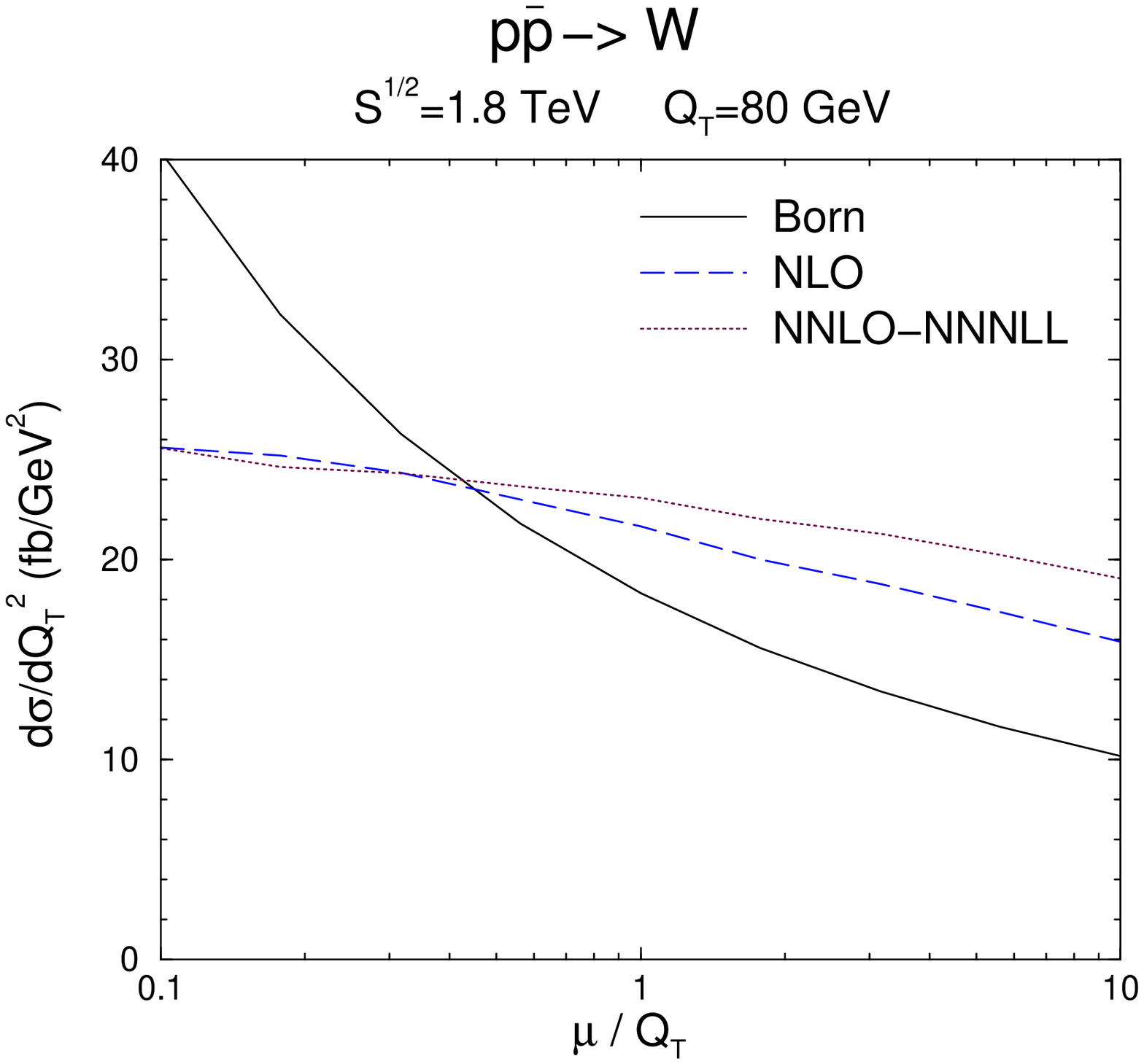,width=8cm}}
\vspace*{8pt}
\caption{$W$-boson production at large $Q_T$ at $\sqrt{S}=1.8$ TeV.}
\end{figure}
   
The $Q_T$ distribution is shown in Fig. 1 at Tevatron Run I,
with $\sqrt{S}=1.8$ TeV. In the left frame we show the
differential distribution $d\sigma/dQ_T^2$ at Born (lowest order), 
NLO, and NNLO, all with scale $\mu=Q_T$,
while in the right frame we show a plot of the scale dependence
at $Q_T=80$ GeV. 
For the NNLO corrections we show both NNLL and NNNLL results.
The NNLL results are complete while in the NNNLL results we have included
the dominant NNNLL terms (more two-loop calculations are needed for an
exact NNNLL calculation \cite{NKtwoloop}).
Throughout we have used the MRST2002 NNLO parton densities \cite{MRST}.
We see that the NNLO corrections
are not very large but they significantly diminish the 
factorization/renormalization scale dependence of the cross section.
 
In Fig. 2 we show similar results for Tevatron Run II, with $\sqrt{S}=1.96$
TeV. In the left frame we plot $d\sigma/dQ_T^2$ with $\mu=Q_T$, while in the 
right frame we show results at $\mu=Q_T/2$ and $2Q_T$.
Again, the reduction of the scale dependence at NNLO is evident:
the two NNLO curves are on top of each other.
Finally, we note that similar results have been derived for 
the related process of direct photon production \cite{NKJO}.
 
\section*{Acknowledgments}

We thank Richard Gonsalves for help with the
NLO corrections. The research of N.K. has been
supported by a Marie Curie Fellowship of the European Community program
``Improving Human Research Potential'' under contract no.
HPMF-CT-2001-01221.
The work of A.S.V. was supported by an Alexander von Humboldt
Postdoctoral Fellowship.

\begin{figure}
\centerline{\psfig{file=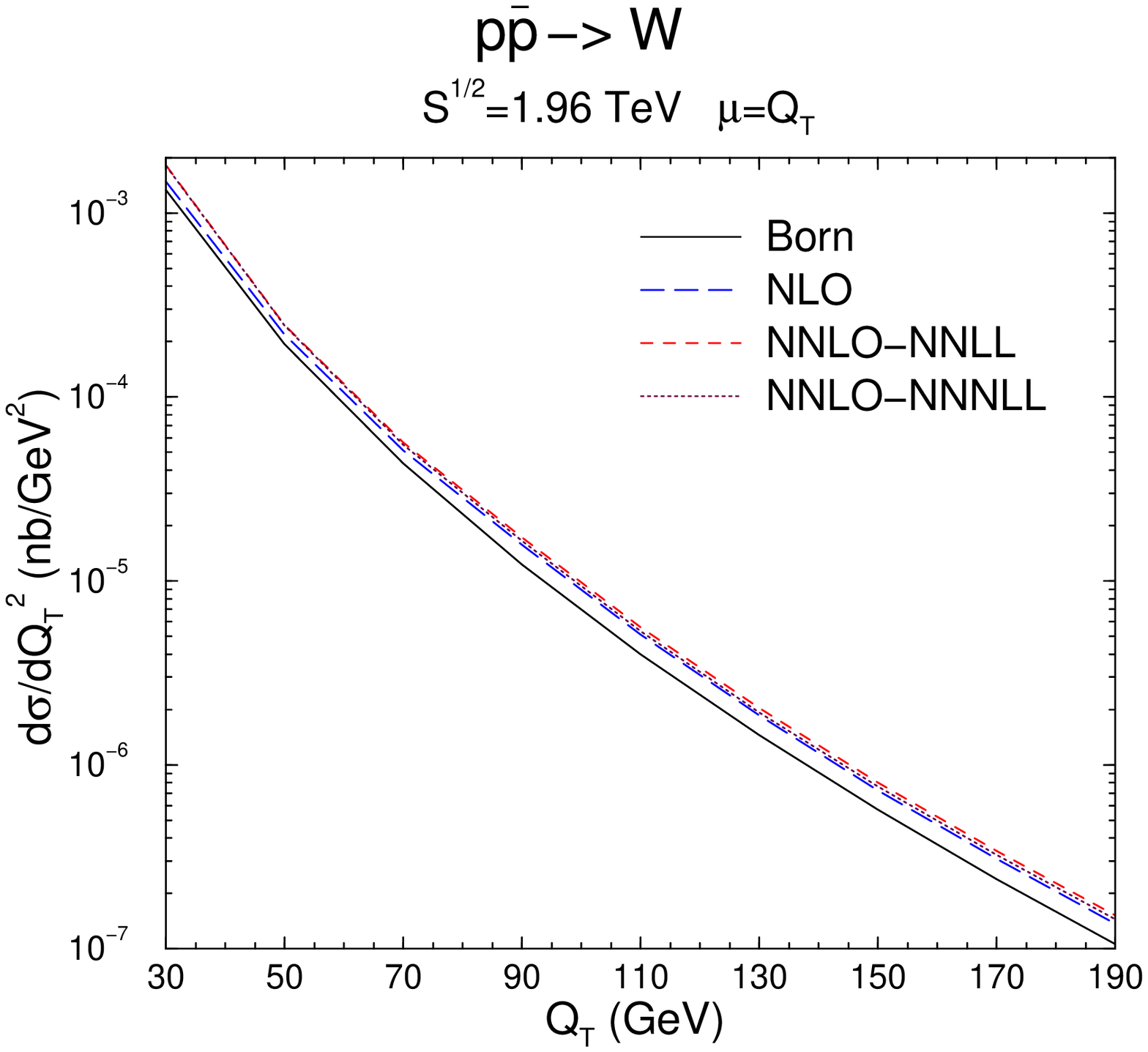,width=8cm} \hspace{5mm}
\psfig{file=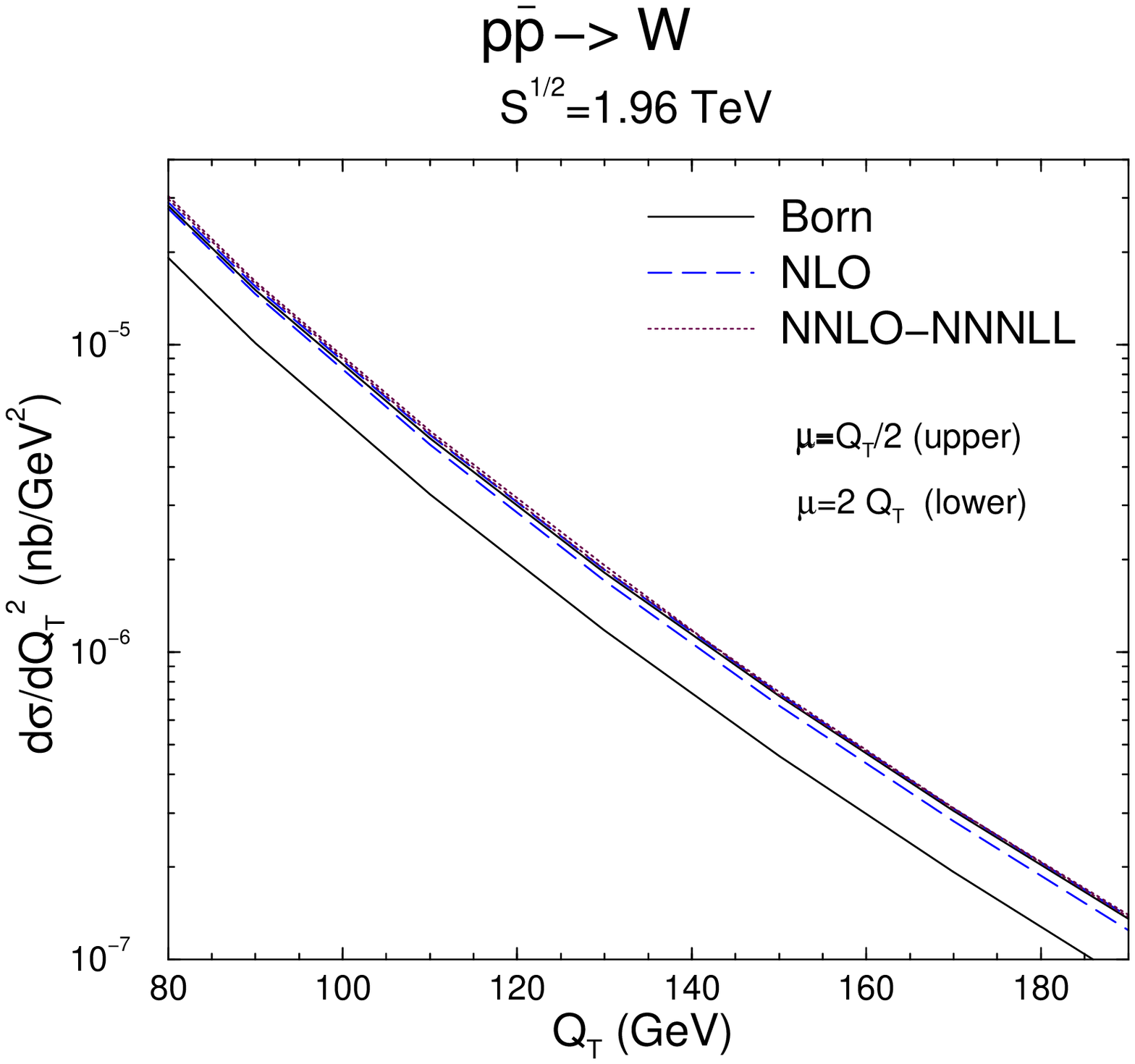,width=8cm}}
\vspace*{8pt}
\caption{$W$-boson production at large $Q_T$ at $\sqrt{S}=1.96$ TeV.}
\end{figure}

\end{document}